\documentclass[onecolumn,showpacs,prd,preprintnumbers,amsmath,amssymb]{revtex4}

\usepackage{graphicx}
\usepackage{dcolumn}
\usepackage{bm}
\usepackage{epsfig}
\usepackage{amssymb}
\usepackage{amsmath}

 \def\tskip{\setlength{\tskip}{5pt}}
\def\colwidth{\setlength{\colwidth}{3.5in}}

\def\prd{Phys. Rev. D~}
\def\pr{Phys. Rep.~}
\def\prl{Phys. Rev. Lett.~}
\def\plb{Phys. Lett. B~}
\def\jcap{JCAP~}
\def\apj{Astrophys. J.~}

\def\apjs{Astrophys. J. Suppl. Ser.~}
\def\aanda{Astron. Astrophys. ~}
\def\jhep{Journal of High Energy Physics ~}
\def\cqg{Classical Quantum Gravity ~}
\def\lnp{Lecture Notes in Physics~}
\def\npb{Nuclear Physics B~}

\newcommand{\lsim}{\mathrel{\hbox{\rlap{\lower.55ex\hbox{$\sim$}} \kern-.3em \raise.4ex \hbox{$<$}}}}
\newcommand{\gsim}{\mathrel{\hbox{\rlap{\lower.55ex\hbox{$\sim$}} \kern-.3em \raise.4ex \hbox{$>$}}}}
\newcommand{\beq}{\begin{equation}}
\newcommand{\eeq}{\end{equation}}
\newcommand{\beqa}{\begin{eqnarray}}
\newcommand{\eeqa}{\end{eqnarray}}
\newcommand{\bea}{\begin{eqnarray}}
\newcommand{\ena}{\end{eqnarray}}

\begin{document}

\title{Testing inflationary consistency relations by the potential CMB observations}

\author{Wen Zhao}
\email{wzhao7@mail.ustc.edu.cn} \affiliation{International Center
for astrophysics, Korea Astronomy and Space Science Institute,
Daejeon, 305-348, Korea}\affiliation{Niels Bohr Institute,
Copenhagen University, Blegdamsvej 17, Copenhagen DK-2100,
Denmark}

\author{Qing-Guo Huang}
\email{huangqg@itp.ac.cn} \affiliation{Key Laboratory of Frontiers
in Theoretical Physics, Institute of Theoretical Physics, Chinese
Academy of Sciences, Beijing 100190, China}

\date{\today}

\begin{abstract}

Testing the so-called consistency relations plays an important
role in distinguishing the different classes of inflation models.
In this paper, we investigate the possible testing for various
single-field inflation models based on the potential future
observations of the cosmic microwave background (CMB) radiation,
including the planned CMBPol mission and the ideal CMB experiment
where only the reduced cosmic weak lensing contamination for the
B-mode polarization is considered. We find that for the canonical
single-field inflation, the phantom inflation and the
potential-driven G-inflation, the consistency relations are quite
hard to be tested: the testing is possible only if $r>0.14$ for
CMBPol mission, and $r>0.06$ for the ideal experiment. However,
the situation could become much more optimistic for the general
Lorentz-invariant single-field inflation model with large non-local
non-Gaussian signal. We find that testing the latter one
class of inflation is possible if $r\gtrsim 10^{-2}$
or even smaller for both CMBPol and ideal CMB experiments.

\end{abstract}


\maketitle

%%%%%%%%%%%%%%%%%%%%%%%%%%%%%%%%%%%%%%%%%%%%%%%%%%%%%%%%%%%%%%%%%%%%%%%%%%%%%%%%%%%%%
%%%%%%%%%%%%%%%%%%%%%%%   SECTION 1     %%   SECTION 1  %%   SECTION 1   %%%%%%%%%%%%%%%%%%%%%%%%%%%%%%%%
%%%%%%%%%%%%%%%%%%%%%%%%%%%%%%%%%%%%%%%%%%%%%%%%%%%%%%%%%%%%%%%%%%%%%%%%%%%%%%%%%%%%%

\section{Introduction \label{section-1}}

Understanding the expansion history of the Universe is a
fundamental task of modern cosmology. The standard hot big-bang
cosmological model is the most successful model to explain various
observations \cite{weinberg2008}. However, in this scenario, one
has to face to the flatness, horizon and monopole puzzles. In
order to solve these problems, various inflation-like scenarios
for the expansion history of the Universe at the very early
time have been proposed \cite{inflation}. The necessity of this
stage can also be understood by the following way: it seems
logical to suggest that our Universe came into being as a
configuration with a Planckian size and a Planckian energy
density, and with a total energy, including gravity, equal to zero
(see \cite{zeldovich1981} and references therein). The newly
created classical configuration cannot reach the average energy
density and size of the presently observed Universe, unless the
configuration experienced a primordial kick, i.e. a inflation-like
stage \cite{zeldovich1981}.

Nowadays there are various inflation models in the market
\cite{inflation}. The problem is how to distinguish these quite
different models from observations. It is well known that the
strong variable gravitational field of the inflationary universe
inevitably generates primordial density perturbations (i.e. scalar
perturbations) and relic gravitational waves (i.e. tensor
perturbations) \cite{grishchuk1974,mukhanov1991}. The former
provides the seed of the large-scale structure formation, and the
latter faithfully encodes the information of the whole expansion
history of the Universe \cite{grishchuk2001}. Searching for the
evidence of these perturbations provides a way to study the
physics in the inflationary stage, and opens an observational
window to explore the physics around the very high energy scale.

The current observations on the cosmic microwave background (CMB)
radiation and the large-scale structure support that the
primordial density perturbations have the nearly scale-invariant
power spectrum which are predicted by the inflation models.
However, it is not sufficient to distinguish various inflation
models, i.e. almost all the models can explain the present
observations, so long as the proper model parameters are chosen.
At present we even have not an obscure picture of this stage: Was
the inflationary stage promoted by a single effective scalar
field, multiple scalar fields or some effective fields with
non-canonical kinetic terms?

Recently, a number of authors have discussed how to test the
inflation models by the current and potential future CMB
observations (see, for instance,
\cite{knox1999,wmap_notation,efstathiou2006,verde2006,cmbpol,peiris2011,easson2010}).
In these works, the authors mainly focused on testing the
canonical single-field slow-roll inflation models, by determining
the scalar spectrum index $n_s$, its running $\alpha_s$ and the
tensor-to-scalar ratio $r$. Different from
these works, in this paper, we shall investigate the possibilities
to confirm or rule out the different single-field inflationary
scenarios (including canonical single-field models, phantom
models, general Lorentz-invariant models and so on) by the
potential future CMB observations.

One of the most powerful tools to distinguish these different
scenarios is to test the so-called consistency relations, which are independent on the inflaton potential and
are quite different in different scenarios. The experimental
determination of the parameters specifying the relic gravitational
waves plays a crucial role in approaching this aim. The CMB has
proved to be a valuable tool in this respect. Relic gravitational
waves leave an observable imprint in the temperature and
polarization anisotropies on the CMB \cite{polnarev1985}, which
provides the unique way to detect the relic gravitational waves
with the largest wavelength.

Although the recent effort, including the WMAP satellite
\cite{komatsu2010,zbg2009}, QUaD \cite{quad}, BICEP \cite{bicep}
and QUIET \cite{currentquiet}, has not found the definite evidence
of relic gravitational waves, their purpose remains one of the key
tasks for the current, upcoming and future CMB observations on the
ground
\cite{quad,bicep,currentquiet,clover,polarbear,quiet,quijote,sptpol,actpol,qubic},
on balloons \cite{ebex,pappa,spider} and in the space
\cite{taskforce,planck,b-pol,litebird,cmbpol,core}.

The accurate measurement of the parameters specifying the relic
gravitational waves, e.g. the spectral index $n_t$, depends on the
full-sky observations of the CMB B-mode polarization field with
the sensitive experiments
\cite{zhao2009a,zhao2009b,ma2010,zhao2011}. The proposed CMBPol
project \cite{cmbpol}, which is taken as a next-generation mission
of Planck satellite \cite{planck}, provides an excellent
opportunity to realize this aim. In this paper, we will carefully
discuss the possibility of testing the consistency relations in
various single-field inflation models by CMBPol mission and an
ideal CMB experiment where only the reduced cosmic weak lensing
contamination for the B-mode polarization is considered. We find
that although it is a very hard task which has been claimed by
many other authors, it is still possible to test these consistency
relations by the CMBPol mission and the ideal CMB experiment, so
long as the amplitude of relic gravitational waves is not too
small. If so, these observations will certainly provide a great
chance for us to investigate the physics in the early Universe and
figure out a natural inflationary scenario.

The outline of the paper is as follows. In Sec.~\ref{section-2},
we introduce how to measure the parameters: the tensor-to-scalar
ratio $r$ and the spectral index $n_t$, by the potential future
CMB observations, including the planned CMBPol mission and the
ideal CMB experiment. Based on these results, in Sec. \ref{sec3},
we carefully discuss the possibility of testing the inflationary
consistency relations in the canonical single-field slow-roll
inflation, the general Lorentz-invariant single-field inflation,
the phantom inflation and
the potential-driven G-inflation. Sec. \ref{sec4} is contributed
as a simple conclusion, which summarizes the main results in this
paper.

\section{Detecting relic gravitational waves in the CMB \label{section-2}}

The main contribution to the observed temperature and polarization
anisotropies of the CMB comes from two types of the cosmological
perturbations, density perturbations and relic gravitational
waves. These perturbations are generally characterized by their
primordial power spectra. These power spectra are usually assumed
to be power-law, which is a generic prediction of a wide range of
scenarios of the early Universe, for example the inflation models
\footnote{In general there might be deviations from a power-law,
which can be parameterized in terms of the running of the spectral
index (see for example \cite{GrishchukSolokhin,turner}), but we
shall not consider this possibility in the current paper.}. Thus,
the power spectra of the perturbation fields take the form
 \bea
 P_s(k)=A_s(k_0)(k/k_0)^{n_s-1},~~~~
 P_t(k)=A_t(k_0)(k/k_0)^{n_t},\label{gw-spectrum}
 \ena
for density perturbations and relic gravitational waves
respectively. In the above expression $k_0$ is an arbitrarily
chosen pivot wavenumber, $n_s$ is the primordial power spectral
index for density perturbations, and $n_t$ is the spectral index
for gravitational waves. $A_s(k_0)$ and $A_t(k_0)$ are the
normalization coefficients determining the absolute values of the
primordial power spectra at the pivot wavenumber $k_0$.

We can also define the tensor-to-scalar ratio as follows
\begin{eqnarray} \label{r-define}
r(k_0)\equiv \frac{A_t(k_0)}{A_s(k_0)},
\end{eqnarray}
which describes the relative contribution of density perturbations
and gravitational waves. The amplitude of gravitational waves
$A_t\left(k_0\right) = r(k_0)A_s(k_0)$ provides us with direct
information on the Hubble parameter in the very early Universe
\cite{GrishchukSolokhin}. More specifically, this amplitude is
directly related to the value of the Hubble parameter $H$ at a time
when perturbation mode with wavenumber $k_0$ crossed the horizon
\cite{wmap_notation}
\begin{eqnarray}
A_t^{1/2} (k_0)= \left.\frac{\sqrt{2}}{M_{\rm
pl}}\frac{H}{\pi}\right|_{k_0/a = H},\label{rh}
\end{eqnarray}
where $M_{\rm pl}=1/\sqrt{8\pi G}$ is the reduced Planck mass. If
we adopt $A_s=2.430\times 10^{-9}$ from the 7-year WMAP
observations \cite{komatsu2010}, the Hubble parameter is
$H\simeq2.67r^{1/2}\times 10^{14}$GeV which only depends on the
value of $r$. In the standard single-field slow-roll inflation
models, the Hubble parameter directly relates to the energy scale
of inflation $V^{1/4}$. The relation (\ref{rh}) follows that
$V^{1/4}\simeq3.35r^{1/4}\times10^{16}$GeV, which has been
emphasized by a number of authors.

Density perturbations and gravitational waves produce temperature
and polarization anisotropies in the CMB, which are characterized
by four angular power spectra $C_{\ell}^{T}$, $C_{\ell}^{C}$,
$C_{\ell}^{E}$ and $C_{\ell}^{B}$ as functions of the multipole
number $\ell$. Here $C_{\ell}^{T}$ is the power spectrum of the
temperature anisotropies, $C_{\ell}^{E}$ and $C_{\ell}^{B}$ are
the power spectra of the so-called E-mode and B-mode of
polarization (note that, density perturbations do not generate
B-mode of polarization \cite{polnarev1985}), and $C_{\ell}^{C}$ is
the power spectrum of the temperature-polarization cross
correlation.

In general, the power spectra $C_{\ell}^{Y}$ (where $Y=T,E,B$ or
$C$) can be presented in the following form
\begin{eqnarray}\label{c-sum}
C_{\ell}^{Y}=C_{\ell}^{Y}({\rm dp})+C_{\ell}^{Y}({\rm gw}),
\end{eqnarray}
where $C_{\ell}^{Y}({\rm dp})$ is the power spectrum due to the
density perturbations, and $C_{\ell}^{Y}({\rm gw})$ is the
spectrum due to gravitational waves.

Since we are primarily interested in the parameters of the
gravitational-wave field, in the following discussion, we shall
work with a fixed cosmological background model. More
specifically, we shall work in the framework of $\Lambda$CDM
model, and keep the background cosmological parameters fixed at
the values determined by a typical model \cite{komatsu2010}
$h=0.705$, $\Omega_b h^2=0.02255$, $\Omega_{m}h^2=0.1126$,
$\Omega_{k}=0$, $\tau_{reion}=0.088$, $A_s=2.430\times 10^{-9}$.
Furthermore, the spectral indices of density perturbations and
gravitational waves are adopted as follows for the simplicity,
\begin{eqnarray}
n_s=1, ~~n_t=0.
\end{eqnarray}
Note that, although the constraint of $n_s$ is quite tight based
on the WMAP observations, its value strongly depends on the
assumption of the primordial power spectrum, i.e. running or no
running (see \cite{komatsu2010} for the details). In this section,
in order to simplify the calculation and without loss of
generality, we assume the scale-invariant power spectra for both
density perturbations and gravitational waves in the fiducial
model.

The CMB power spectra $C_{\ell}^{Y}$ are theoretical constructions
determined by ensemble averages over all possible realizations of
the underlying random process. However, in real CMB observations,
we only have access to a single sky, and hence to a single
realization. In order to obtain information on the power spectra
from a single realization, it is required to construct estimators
of power spectra. In order to differentiate the estimators from
the actual power spectra, we shall use the notation $D_{\ell}^Y$
to denote the estimators while retaining the notation $C_{\ell}^Y$
to denote the power spectrum. The probability distribution
functions for the estimators predict the expectation values of the
estimators
\begin{equation} \langle D_{\ell}^Y\rangle=C_{\ell}^Y,\label{mean}
\end{equation}
and the standard deviations
\begin{eqnarray}
(\sigma_{D_{\ell}^X})^2&=&\frac{2(C_{\ell}^X+N_{\ell}^X)^2}{(2\ell+1)f_{\rm sky}},~~(X=T,E,B)\nonumber\\
(\sigma_{D_{\ell}^C})^2&=&\frac{(C_{\ell}^T+N_{\ell}^T)(C_{\ell}^E+N_{\ell}^E)+(C_{\ell}^C+N_{\ell}^C)^2}{(2\ell+1)f_{\rm
sky}},\label{variance}
\end{eqnarray}
where $f_{\rm sky}$ is the sky-cut factor, and $N_{\ell}^Y$ are
the noise power spectra, which are all determined by the specific
experiments.

In order to estimate the parameters $r$ and $n_t$ characterizing
the gravitational-wave background, we shall use an analysis based
on the likelihood function \cite{cosmomc}. In previous work
\cite{zhao2009a}, we have analytically discussed how to constrain
the parameters of the relic gravitational waves, $r$ and $n_t$, by
the CMB observations. We found that in general, the constraints on
$r$ and $n_t$ correlate with each other. However, if we consider
the tensor-to-scalar ratio at the best-pivot wavenumber $k_t$, the
constraints on $r$ and $n_t$ become independent on each other, and
the uncertainties $\Delta{r}$ and $\Delta{n_t}$ have the minimum
values. We have derived the analytical formulae to calculate the
quantities: the best-pivot wavenumber $k_t$, and the uncertainties
of the parameters $\Delta r$ and $\Delta {n_t}$, which provides a
simple and quick method to investigate the detection abilities of
the future CMB observations. In \cite{zhao2009a}, we have also
found that these analytically results are well consistent with the
simulation results by using the Markov-Chain Monte-Carlo method.
We shall briefly introduce these results in this section.

It is convenient to define the quantities as follows
 \bea
 a_{\ell}^{Y}\equiv \frac{C_{\ell}^Y({\rm gw})}{\sigma_{D_{\ell}^Y}},~~b_{\ell}\equiv
 \ln\left(\frac{\ell}{\ell_t}\right),~~d_{\ell}^Y\equiv
 \frac{D_{\ell}^Y-C_{\ell}^Y({\rm dp})}{\sigma_{D_{\ell}^Y}},
 \ena
where $\sigma_{D_{\ell}^Y}$ is the standard deviation of the
estimator $D_{\ell}^Y$, which can be calculated by
Eq.(\ref{variance}). We should notice that the quantity
$d_{\ell}^Y$ is dependent of random date $D_{\ell}^Y$. By
considering the relations in (\ref{mean}) and (\ref{c-sum}), we
can obtain that $\langle d_{\ell}^Y\rangle=a_{\ell}^Y$, which
shows that $d_{\ell}^Y$ is an unbiased estimator of $a_{\ell}^Y$.
$\ell_t$ is the so-called best-pivot multipole, which is
determined by solving the following equation \cite{zhao2009a}:
 \bea\label{ltstar}
 \sum_{\ell}\sum_{Y}a_{\ell}^{Y2}b_{\ell}=0.
 \ena
So the value of $\ell_t$ depends on the cosmological model, the
amplitude of gravitational waves, and noise power spectra by the
quantity $a_{\ell}^Y$. The best-pivot wavenumber $k_t$ relates to
$\ell_t$ by the approximation relation \cite{zhao2009a},
\begin{equation}\label{kl}
k_t\simeq \ell_t\times 10^{-4}{\rm Mpc}^{-1}.
\end{equation}

Note that, in our previous work \cite{zhao2009a}, the best-pivot
wavenumber $k_t$ and multipole $\ell_t$ have been denoted as
$k_t^*$ and $\ell_t^*$, respectively. Once the value of $\ell_t$
is obtained, the uncertainties $\Delta r$ and $\Delta n_t$ can be
calculated by the following simple formulae
 \bea\label{uncertainties}
 \Delta r=r/\sqrt{\sum_{\ell}\sum_{Y}a_{\ell}^{Y2}},~~~~
 \Delta
 n_t=1/\sqrt{\sum_{\ell}\sum_{Y}(a_{\ell}^{Y}b_{\ell})^2}.
 \ena
As usual, we can define the signal-to-noise ratio $S/N\equiv
r/\Delta r$. Using (\ref{uncertainties}), we get
 \bea\label{snr}
 S/N=\sqrt{\sum_{\ell}\sum_{Y}a_{\ell}^{Y2}}.
 \ena
Here, we mention that in Eq. (\ref{uncertainties}) and throughout
the paper below, the quantity $r$ denotes the tensor-to-scalar
ratio at the best pivot-wavenumber, i.e. $r\equiv r(k_t)$, which
has been written as $r^*$ in the previous work \cite{zhao2009a}.

For a given gravitational-wave background, the values of $S/N$ and
$\Delta n_t$ mainly depend on two experimental quantities: the
total noise level of the experiment and the surveyed sky area. The
lower noise and larger sky survey follow a larger $S/N$ and a
smaller $\Delta n_t$. In the previous works
\cite{zbg2009,zhao2009a,zhao2009b,ma2010,zhao2011}, we have
carefully investigated the detection abilities of various future
CMB experiments. We found that in the optimistic case, the
launched Planck satellite is expected to find the signal of
gravitational waves if $r>0.03$, consistent with
\cite{efstathiou2009}. The ground-based experiments, such as QUIET
and POLARBEAR, are expected to have a detection if $r>0.01$.
However, both of them cannot well determine the spectral index
$n_t$, due to the large noise level for Planck satellite and the
small sky-cut factor for the ground-based experiments. Even if
combining the Planck and ground-based POLARBEAR experiments, we
only obtain $\Delta n_t=0.1$ for the case with the
tensor-to-scalar ratio $r=0.1$ \cite{zhao2009b}, which is not
accurate enough to distinguish different inflation models.

It has been noticed that the well detection of gravitational waves
needs a full-sky observation by the high-sensitivity detectors.
The proposed CMBPol mission provides an excellent opportunity in
this respect \cite{cmbpol}. CMBPol mission is expected to have a
full-sky survey for the CMB temperature and polarization fields,
and the instrumental noises are close to, or even lower than, the
cosmic weak leansing contamination for the B-mode polarization.
So, in this paper, we discuss the detection of relic gravitational
waves and the distinguishing of the inflation models by the
potential CMBPol observations. For the CMBPol mission, we consider
a sky-cut factor $f_{\rm sky}=0.8$, proposed in the CMBPol white
book \cite{cmbpol}. The total noises of CMBPol observation mainly
include three parts: the instrumental noises, the foreground
contaminations (including the synchrotron and dust emissions) and
cosmic weak lensing contamination for the B-mode polarization. For
the instrumental noises, there are several proposals
\cite{cmbpol,cmbpol2}. In this paper, we shall focus on the
middle-cost EPIC-2m proposal. The detailed calculation of the
total noise power spectra of EPIC-2m is given in our recent work
\cite{zhao2011}, where the analytical formulae are given to
calculate the noise power spectrum. Throughout this paper, we have
used the CAMB package \cite{camb} to calculate the CMB power
spectra, and the contaminations due to the cosmic weak lensing. We
notice that, the total noise power spectra depend on the assumed
parameters $\sigma^{\rm fore}$ and $\sigma^{\rm lens}$, which
describe the residual fractions of foreground emissions and lensed
B-mode polarization considered as the effective noises. In this
paper, we shall focus on the optimistic case with the assumed
parameters ($\sigma^{\rm fore}$, $\sigma^{\rm lens}$) being
($0.01$, $0.5$). Note that, in our discussion, we have not
considered the leakage from the E-mode into the B-mode
polarization due to the partial sky analysis. we assume this E-B
mixture can be properly avoided (or deeply reduced) by
constructing the pure E-mode and B-mode polarization fields
\cite{ebmixture}.

%The instrumental noises of this mission can be found in
%\cite{cmbpol} (Appendix C). For the foreground contaminations, we
%consider the \emph{Dust B} model in \cite{cmbpol} and the
%optimistic foreground removal with $\sigma^{\rm fg}=0.01$, i.e.
%only $1\%$ reduced foreground emissions are considered as the
%noise. The B-mode polarization generated by weak lensing is
%considered as another kind of effective noise. It was claimed that
%for the observation with experiment like EPIC-2m, one expects to
%delense half of the total lensed B-mode power spectrum
%\cite{cmbpol-delense}. So in the discussion, we consider the
%reduced lensed B-mode polarization with the residual factor
%$\sigma^{\rm lens}=0.5$ as the effective noise. The detailed
%calculation of the noise power spectra can be found in
%\cite{zhao2011}, which will not be shown here. Note that, in our
%discussion, we have not considered the leakage from the E-mode
%into the B-mode polarization due to the partial sky analysis. we
%assume this E-B mixture can be properly avoided (or deeply
%reduced) by constructing the pure E-mode and B-mode polarization
%fields \cite{ebmixture}.

\begin{figure*}[t]
\begin{center}
\includegraphics[width=17cm,height=10cm]{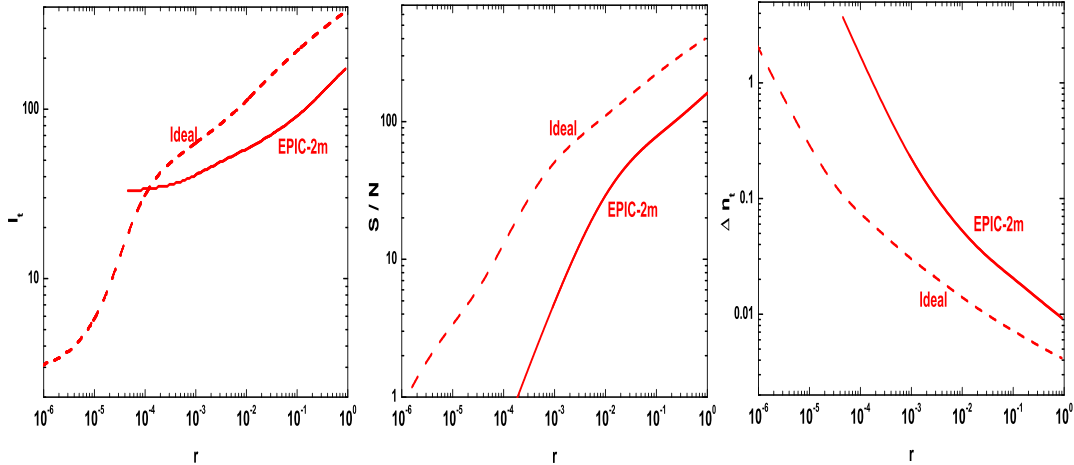}
\end{center}
\caption{The figures show the values of the best-pivot multipole
$\ell_t$ (left panel), signal-to-noise ratio $S/N$ (middle panel)
and the uncertainty of the spectral index $\Delta n_t$ (right
panel) as functions of the tensor-to-scalar ratio
$r$.}\label{figure1}
\end{figure*}

Taking into account the total noises, and using the formulae in
Eqs. (\ref{ltstar}), (\ref{uncertainties}), (\ref{snr}), we
calculate the quantities $\ell_t$, $S/N$ and $\Delta n_t$ as
functions of the input tensor-to-scalar ratio $r$. The results are
shown in Fig. \ref{figure1} (the red solid lines), which are
consistent with the previous works \cite{zhao2011,ma2010,cmbpol}.
Fig. \ref{figure1} shows that as expected, the model with a larger
$r$ follows a larger $S/N$, and a smaller $\Delta n_t$. It is
interesting to find that when $r>0.001$, EPIC-2m can detect the
signal of gravitational waves at more than $5\sigma$ level. For
$r=0.01$ case, we have $S/N=29$, and for $r=0.1$, we have
$S/N=78$. So we conclude that EPIC-2m can well detect the signal
of gravitational waves so long as the tensor-to-scalar ratio is
larger than $0.001$. At the same time, the determination of the
spectral index $n_t$ is also quite accurate. If $r=0.001$, one has
$\Delta n_t=0.2$, and if $r=0.01$, one has $\Delta n_t=0.05$.
Especially, for $r=0.1$, the uncertainty reduces to $\Delta
n_t=0.02$. So, it would be a quite powerful tool to study the
physics in the early Universe, especially for distinguishing
different inflation models, which will be shown in Sec.
\ref{sec3}.

%%%%%%%%%%%%%%%%%%%%%%%%%%%%%%%%%%%%%%%%%  figure 1, figure 1, figure 1%%%%%
%%%%%%%%%%%%%%%%%%%%%%%%%%%%%%%%%%%%%%%%%%%%%
%%%%%%%%%%%%%%%%%%%%%%%%%%%%%%%%%%%%%%%%%  figure 1, figure 1, figure 1%%%
%%%%%%%%%%%%%%%%%%%%%%%%%%%%%%%%%%%%%%%%%%%%%%%

It was noticed that the detection abilities of the future CMB
experiments for the relic gravitational waves are limited by the
cosmic variance and cosmic weak lensing effect (see
\cite{lensingreview} and references therein). Especially, when the
instrumental noise power spectra of the future experiment become
smaller than $\sim 10^{-6}\mu$K$^2$, the weak lensing
contamination for the B-mode polarization could be dominant among
the total noises, and forms a detection limit for the CMB
experiments \cite{song2002,zhao2009a}. A number of works have
discussed methods to subtract the lensed B-mode signal (see
\cite{lensing1,lensing2}). In \cite{lensing2}, the authors claimed
that a reduction in lensing power by a faction of $40$, i.e. a
residual faction $\sigma^{\rm lens}=0.025$, is possible using an
approximate iterative maximum-likelihood method. For this reason,
as an idealized scenario, we shall also consider the case with
reduced cosmic lensing noise with $\sigma^{\rm lens}$. In this
ideal case, we assume an exactly full sky survey with $f_{\rm
sky}=1$. We also assume that there are no instrumental noises or
foreground emissions.

By the similar steps, we calculate the quantities $\ell_t$, $S/N$
and $\Delta n_t$ as functions of the input tensor-to-scalar ratio
$r$ in this ideal case. The results are also shown in Fig.
\ref{figure1} with dashed lines, consistent with \cite{zhao2009a}.
We find that the gravitational waves with $r>3.7\times 10^{-6}$
can be detected at more than $2\sigma$ level. This can be treated
as the detection limit of the CMB experiments. This lower limit
corresponds to the Hubble parameter $H\simeq 3.1\times
10^{11}$GeV, and the energy scale of inflation $V^{1/4}\simeq
1.5\times 10^{15}$GeV. From Fig. \ref{figure1}, we also find that
when $r>2\times10^{-5}$, the signal-to-noise ratio becomes quite
large, i.e. $S/N>5$. In this idealized situation, the uncertainty
of the spectral index $n_t$ also becomes very small. We can
constrain $n_t$ to the level $\Delta n_t=0.014$ if $r=0.01$, and
$\Delta n_t=0.007$ for $r=0.1$. Distinguishing different inflation
models in this ideal case will also be presented in Sec.
\ref{sec3}.

\section{Testing the inflationary consistency relations
\label{sec3}}

Nowadays the inflationary scenario has been widely accepted by
almost all of the cosmologists. Inflation can naturally explain
the well observed primordial density perturbations with a nearly
scale-invariant power spectrum. However, we still do not know
which inflation model describes our real Universe, since almost
all the models are compatible with the current observations if the
proper model parameters are adopted. So distinguishing different
inflation models, especially different classes of models, is the
key task for the future research.

The so-called consistency relation gives a clear difference for
different classes of inflation. For example, for the cannonical
single-field slow-roll inflation models, the consistency relation
$r=-8n_t$ is held, which provides a unique way to confirm or rule
out this class of models. In this section, we shall carefully
investigate the possibility of testing the consistency relations
for the canonical single-field slow-roll inflation, the general
Lorentz-invariant single-field inflation, the phantom inflation and the potential-driven
G-inflation, by the potential future observations, such as the planned
CMBPol mission and the ideal CMB experiment.

\subsection{Canonical single-field slow-roll inflation model \label{sec3.1}}

First of all, we focus on the simplest version of inflation, i.e.
the canonical single-field slow-roll inflation model. In this
scenario, the dynamics of Universe is governed by a scalar field
(the inflaton) $\phi$ with canonical kinetic term. The inflaton
slowly rolled down its flat potential in the inflationary
stage. Inflation ended when the slow-roll conditions were broken
down, and the inflaton decayed into the relativistic particles and
re-heated the Universe. See some recent discussions in \cite{Alabidi:2008ej}.

The Lagrangian for the canonical single-field inflation model is
given by \cite{inflation}
 \bea
 \mathcal{L}(\phi)=\frac{1}{2}(\partial \phi)^2
 -V(\phi), \label{single-field}
 \ena
where $V(\phi)$ is the potential of inflaton field $\phi$.
Different $V(\phi)$ describes the different inflation models. In
the inflationary stage, the potential energy of the inflaton
dominates over its kinetic energy, and $V(\phi)$ should be quite
flat. Thus, we can define the slow-roll parameter
 \bea
 \epsilon=-\frac{\dot{H}}{H^2}\simeq\frac{M_{\rm pl}^2}{2}\left(\frac{V'}{V}\right)^2,
 \ena
where the \emph{dot} denotes $d/dt$, and the \emph{prime} denotes
$d/d\phi$. This parameter $\epsilon$ should be much smaller than one
during inflation.

The tensor-to-scalar ratio $r$ and the tensor spectral index $n_t$
in this scenario are related to the slow-roll parameter $\epsilon$
by \cite{inflation}
 \bea
 r=16\epsilon, ~ n_t=-2\epsilon.
 \ena
The above equations lead to the so-called consistency relation for
the canonical single-field slow-roll inflation \cite{inflation} (For a
detailed critical discussion of this consistency relation see the
last paper in \cite{grishchuk1974}),
 \bea
 n_t=-r/8. \label{consistency1}
 \ena
This consistency relation is independent on the form of the
potential and valid for all the single-field slow-roll inflation
models with canonical kinetic terms \footnote{It was pointed out
that this consistency relation could be violated if considering
the trans-Planckian physics in the early Universe
\cite{transplanck}. So in this sense, the testing of this
consistency relation also provides the change to study the
trans-Planckian physics.}. So testing this relation provides a
model-independent criteria to confirm or rule out the canonical
single-field slow-roll inflation models.

From (\ref{consistency1}), we find that testing this relation
depends on the measurement of the gravitational-wave's parameters
$r$ and $n_t$. Since the absolute value of $n_t$ is expected to be
one order smaller than that of $r$ (see Eq. (\ref{consistency1})),
and the measurement of $n_t$ is much more difficult than $r$
\cite{zhao2009a,zhao2009b,ma2010}. How well we can measure the
spectral index $n_t$ plays a crucial role in testing the
consistency relation (\ref{consistency1}).

Now, let us discuss how well this consistency relation can be
tested by the potential future CMB observations, which has been
partly discussion in the previous works
\cite{smith2006}\cite{ma2010,zhao2011}\cite{easson2010}. Here, we
will revisit this problem based on the discussion in Sec.
\ref{section-2}, where a best-pivot wavenumber number and the best
determinations of the parameters $r$ and $n_t$ are considered.

\begin{figure}[t]
\begin{center}
\includegraphics[width=13cm,height=10cm]{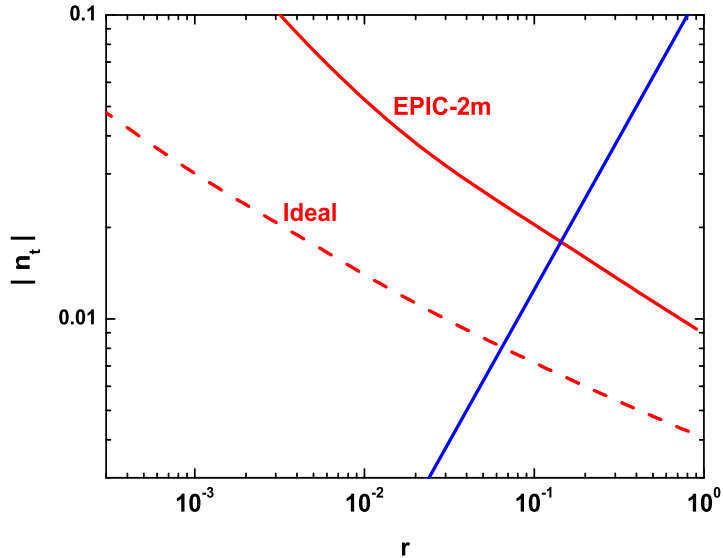}
\end{center}
\caption{For EPIC-2m mission and the ideal CMB experiment, the
value of $\Delta n_t$ compares with that of $|n_t|=r/8$ (blue
solid line). Note that, the red lines are identical to those in
the Fig. \ref{figure1} (right panel). }\label{figure2}
\end{figure}

The uncertainty $\Delta n_t$ as a function of the input $r$ is
replotted in Fig. \ref{figure2} (red lines). We investigate if the
future experiments might distinguish the tilted gravitation waves
from the scale-invariant one (i.e. $n_t=0$). To attain the goal of
this testing, in Fig. \ref{figure2} we compare the value of
$|n_t|=r/8$ with that of $\Delta n_t$ \footnote{In principle, for
any given $n_t$, we should compare its value with that of $\Delta
n_t$, derived from the model with this $n_t$ as input. However, we
notice that in the inflation models considered in this paper, the
value of $n_t$ is always very close to zero, which follows that
the difference between these $\Delta n_t$'s and that in Sec.
\ref{section-2} are very small. Throughout this paper, we shall
ignore this little difference.}. If $\Delta n_t < |n_t|$, then the
constraint on $n_t$ is tight enough to allow the consistency
relation to be tested. We find that $\Delta n_t < |n_t|$ is
satisfied only if $r>0.14$ for the EPIC-2m experiment, which is
quite close to the current upper limit of $r$ obtained from the
7-year WMAP observations \cite{komatsu2010}. So we conclude that
there is only a small space left for testing the consistency
relation for CMBPol. This is the reason why many people claimed
that the consistency relation is difficult to test \cite{cmbpol}.
However, from Fig. \ref{figure2}, we find that this situation can
be slightly alleviated for the ideal observation. In this case,
$\Delta n_t < |n_t|$ is satisfied so long as $r>0.06$. Thus, it is
possible to test the consistency relation (\ref{consistency1})
only for some inflation models with fairly large $r$.

Let us consider two specific models. First, we discuss the
\emph{chaotic inflation}. The prototype for chaotic inflation
involves a single polynomial term $V(\phi)=\Lambda_p(\phi/\mu)^p$
with $p>0$ \cite{liddle1983}. Here, the scale $\mu<M_{\rm pl}$ is
relevant for the higher-dimensional terms in this effective
potential. The chaotic inflation models of this form make the
following prediction \cite{cmbpol}
 \bea
 r=8\left(\frac{p}{p+2}\right)(1-n_s).
 \ena
If assuming $n_s=0.968$ \cite{komatsu2010} and $p=2$, we have
$r=0.128$. So in this model, the consistency relation could be
tested by the ideal experiment. Another typical model we consider
is the \emph{hill-top models} with quadratic term, which has the
potential form
$V(\phi)=V_0\left[1-\left({\phi}/{\mu}\right)^p\right]$ with $p\ge
2$ and $\phi<\mu$. This potential is considered as an
approximation to a generic symmetry-breaking potential
\cite{inflation,cmbpol}. If $p=2$, the value of $r$ can be
expressed as \cite{cmbpol}
 \bea
 r=8(1-n_s)\exp[-1-N_e(1-n_s)],
 \ena
where $N_e$ is the number of e-folds, taking to be in the range
$N_e\in[40,~70]$ based on the current observations of the CMB
\cite{komatsu2010}. For $n_s=0.968$, we find that $r\in
[0.010,~0.026]$. So, if this model describes our real Universe,
the consistency relation cannot be tested even in the ideal case.

%%%%%%%%%%%%%%%%%%%%%%%%%%%%%%%%%%%%%%%%%  figure 1, figure 1, figure 1%%%%%
%%%%%%%%%%%%%%%%%%%%%%%%%%%%%%%%%%%%%%%%%%%%%
%%%%%%%%%%%%%%%%%%%%%%%%%%%%%%%%%%%%%%%%%  figure 1, figure 1, figure 1%%%
%%%%%%%%%%%%%%%%%%%%%%%%%%%%%%%%%%%%%%%%%%%%%%%

\subsection{General Lorentz-invariant single-field inflation model \label{sec3.2}}

In this subsection, we will consider the general Lorentz-invariant
single-field inflation model, in particular the model with
non-canonical kinetic terms, such as K-inflation
\cite{kinflation}, Dirac-Born-Infeld (DBI) inflation
\cite{alishahiha2004}, the power-law kinetic inflation
\cite{seery2005,panotopoulos2007}
and so on. These models may be motivated by the high-dimensional
superstring theory or brane theory \cite{cmbpol}.

The Lagrangian for general Lorentz-invariant single-field
inflation takes the form
 \bea
 \mathcal{L}(\phi)= P(X,\phi), \label{kinflation}
 \ena
where $X\equiv(\partial \phi)^2/2$. Canonical single-field
inflation is included in (\ref{kinflation}). The function
$P(X,\phi)$ corresponds to the pressure of the scalar fluid, while
the energy density is $E=2XP_{,X}-P$, where $P_{,X}\equiv
\partial P/\partial X$. An important parameter in this model is
the speed of sound, which is defined by
 \bea
 c_s^2\equiv\frac{P_{,X}}{E_{,X}}=\frac{P_{,X}}{P_{,X}+2XP_{,XX}}.
 \ena
For the canonical single-field inflation we have $c_s^2=1$.
However, in the general K-inflation model, the value of $c_s^2$
can be larger or smaller than speed of the light
\cite{kinflation}.

The parameters related to the gravitational waves can be
quantified in terms of the sound speed $c_s$ and the slow-roll
parameter $\epsilon\equiv-\dot{H}/H^2$ \cite{kinflation}, i.e.
 \bea
 r=16\epsilon c_s,~~n_t=-2\epsilon.
 \ena
Thus, the consistency relation becomes
 \bea
 n_t=-\frac{r}{8c_s}. \label{consistency2}
 \ena
When $c_s=1$, Eq.~(\ref{consistency1}) is naturally recovered.
However, when $c_s\gg1$ or $c_s\ll1$, the difference between
(\ref{consistency2}) and (\ref{consistency1}) becomes obvious, and
makes it possible to distinguish these two classes of models.

Due to the dependence of $c_s$ in the consistency relation
(\ref{consistency2}), it is necessary to consider other
observation to constrain the sound speed in the models. In
addition to directly constrain $c_s$, there are some other
possibility to constrain the sound speed by the observations. For
the models with $c_s\neq 1$, higher derivative terms in the
Lagrangian are included and usually a large non-local form
bispectrum is predicted. However the full bispectrum is controlled
by two parameters \cite{chen2007}: $c_s^2$ and
$\lambda/\Sigma=(X^2P_{,XX}+{2\over
3}X^3P_{,XXX})/(XP_{,X}+2X^2P_{,XX})$. Therefore we need two
observables to fix $c_s^2$. Fortunately, two independent templates
have been well defined for measuring the non-local form
bispectrum: equilateral and orthogonal forms whose sizes are
respectively measured by two non-Gaussian parameters $f_{\rm
NL}^{\rm equil}$ and $f_{\rm NL}^{\rm orth}$
\cite{komatsu2000,Babich2004,Senatore2009,komatsu2010}. In some
specific cases, such as the DBI inflations \cite{alishahiha2004}
and power-law kinetic inflations
\cite{seery2005,panotopoulos2007}, $c_s^2$ and $\lambda/\Sigma$
can be determined by the unique parameter $f_{\rm NL}^{\rm
equil}$. However, for the general Lorentz-invariant single-field
inflations, in \cite{other2} one of us (QGH) found that the speed
of sound $c_s$ can be fixed once both $f_{\rm NL}^{\rm equil}$ and
$f_{\rm NL}^{\rm orth}$ are detected, i.e.
 \bea
 \frac{1}{c_s^2}-1=-1.260 f_{\rm NL}^{\rm equil}-23.19 f_{\rm NL}^{\rm
 orth}.
 \ena

The constraints on the non-Gaussian parameters by the current and
potential future CMB observations have been discussed by a number
of authors
\cite{komatsu2000,komatsu2010,zaldarriaga2004,yadav2007,smith2009a}.
Different from the constraint on the relic gravitational waves,
the detection of non-Gaussian signal mainly depends on the
observations of CMB $TT$, $TE$ and $EE$ power spectra. So, we
expect that there is no degeneration between constraints of the
non-Gaussian parameters and those of the gravitational-wave's
parameters. The current 7-year WMAP data indicate a constraint on
the non-Gaussian parameters as follows \cite{komatsu2010},
 \beqa
 -214<f_{\rm NL}^{\rm equil}<266, ~~{\rm and}~~
 -410<f_{\rm NL}^{\rm orth}<6~(95\%{\rm ~C.L.}).
 \eeqa
These constraints have been obtained using the temperature signal
only. The upcoming Planck satellite will improve this to the level
$\Delta f_{\rm NL}^{\rm equil}\simeq\Delta f_{\rm NL}^{\rm
orth}\simeq25$, mainly by the observations of the E-mode
polarization signal. In addition, a satellite mission such as
CMBPol, dedicated to polarization and cosmic variance limited up
to $\ell\sim 2000$, would be able to further improve on Planck by
a factor of order 1.6, reaching $\Delta f_{\rm NL}^{\rm
equil}\simeq\Delta f_{\rm NL}^{\rm orth}\simeq14$ \cite{cmbpol,komatsu}.
This  can also be treated as the detection ability of the ideal
CMB experiment, due to the cosmic variance limit.

\begin{figure}[t]
\begin{center}
\includegraphics[width=13cm,height=10cm]{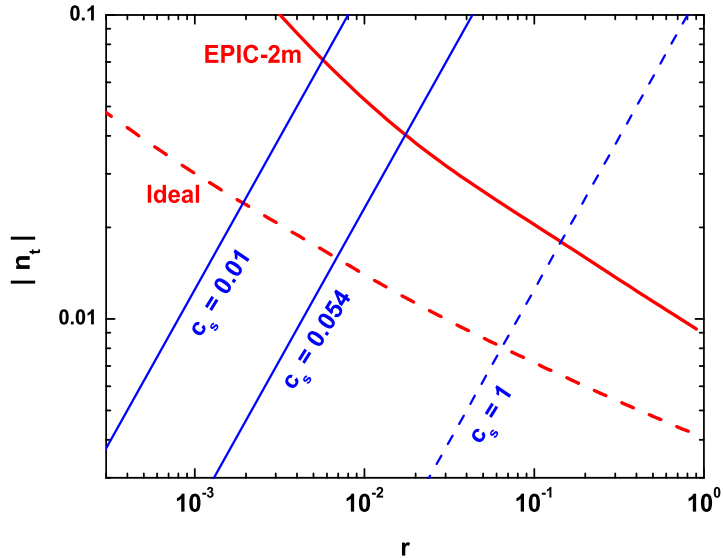}
\end{center}
\caption{For EPIC-2m mission and the ideal CMB experiment, the
value of $\Delta n_t$ compares with that of $|n_t|=r/(8c_s)$ (blue
lines). Note that, the red lines are identical to those in the
Fig. \ref{figure1} (right panel), while the dashed blue line is
identical to the blue line in Fig. \ref{figure2}. }\label{figure3}
\end{figure}

In the limit of $f_{\rm NL}^{\rm equil}=-214$ and $f_{\rm NL}^{\rm
orth}=-410$, signal of non-Gaussianity will be quite well observed
by the future observations, such as Planck satellite
\cite{planck}. In this limit, we have $c_s=0.01$ and the
consistency relation becomes $n_t=-12.5 r$. In Fig. \ref{figure3},
we compare values of $|n_t|=12.5r$ (blue solid line) with $\Delta
n_t$. We find that $\Delta n_t<|n_t|$ is satisfied if $r>0.006$
for EPIC-2m experiment, and $r>0.002$ for the ideal experiment.
Comparing with the conclusion in Sec.\ref{sec3.1}, the quite
promising results are expected for the study of inflation.

%%%%%%%%%%%%%%%%%%%%%%%%%%%%%%%%%%%%%%%%%  figure 1, figure 1, figure 1%%%%%
%%%%%%%%%%%%%%%%%%%%%%%%%%%%%%%%%%%%%%%%%%%%%
%%%%%%%%%%%%%%%%%%%%%%%%%%%%%%%%%%%%%%%%%  figure 1, figure 1, figure 1%%%
%%%%%%%%%%%%%%%%%%%%%%%%%%%%%%%%%%%%%%%%%%%%%%%

\begin{figure}[t]
\begin{center}
\includegraphics[width=13cm,height=10cm]{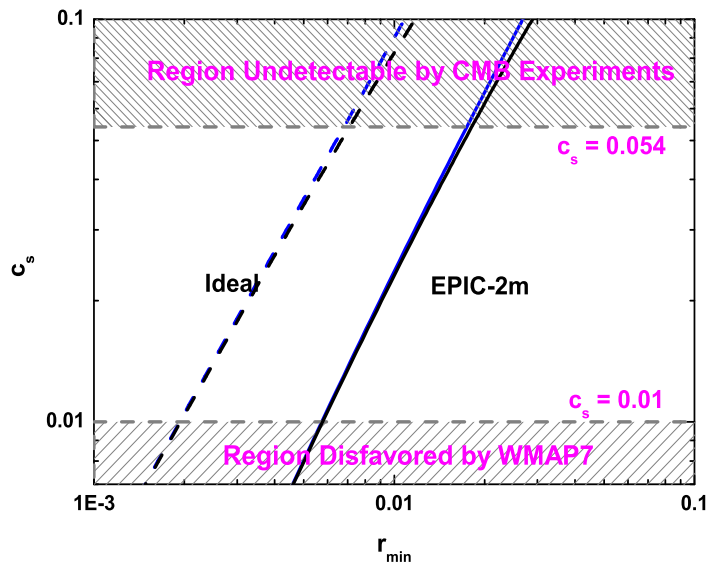}
\end{center}
\caption{The values of $r_{\rm min}$ (blue lines) and $r'_{\rm
min}$ (black lines) as functions of the speed parameter $c_s$ for
both EPIC-2m mission and the ideal CMB experiment.
}\label{figure4}
\end{figure}

However, we have to mention that this is a too optimistic case.
Actually, the values of $|f_{\rm NL}|$ could be much smaller. With
the decreasing of $|f_{\rm NL}|$, the value of $c_s$ increases,
and the testing of consistency relation becomes more and more
difficult. A non-Gaussian signal exceeding $|f_{\rm NL}|\ge 14$
will be detectable from the future CMB observations \cite{cmbpol}.
So as another limit case, let us assume $f_{\rm NL}^{\rm
equil}=f_{\rm NL}^{\rm orth}=-14$, which follows that $c_s=0.054$
and the consistency relation $n_t=-2.31r$. From Fig.
\ref{figure3}, we find that $\Delta n_t<|n_t|$ is satisfied if
$r>0.017$ for EPIC-2m experiment, and $r>0.007$ for the ideal
experiment. We mention that when $|f_{\rm NL}^{\rm equil}|<14$ or
$|f_{\rm NL}^{\rm orth}|<14$, the detection of non-Gaussian signal
becomes impossible, and testing of the consistency relation by
this method becomes impossible as well. In Fig. \ref{figure4}, we
plot the value of $r_{\rm min}$ (where $|n_t|=\Delta n_t$ is
satisfied) for different values of $c_s$ in blue lines. In this
figure, we use the dashed grey lines to label two limit cases
considered above. It is important to mention that these results
are only correct for the non-canonical Lorentz-invariant
single-field inflation models. As a subclass of the general
single-field inflation model, the canonical inflation model has a
definite prediction $c_s=1$. So the sound speed is not a free
parameter for this subclass of models, and we do not need to
constrain $c_s$ for the testing of the consistency relation, which
has been clearly discussed in Section \ref{sec3.1}

Recall that an important goal of inflation programs is to test
whether inflation really arises from the canonical single-field
slow-roll models. To attain the goal of this testing, we should
compare $\Delta n_t$ with the quantity $\delta n_t \equiv
|n_t-r/8|$. If $\Delta n_t < \delta n_t$, the constraint of $n_t$
is tight enough to allow this test. Similar to above, for any
given $c_s$, we define $r'_{\rm min}$ (where $\delta n_t=\Delta
n_t$ is satisfied). This is the minimal $r$, when the test is
allowed. In Fig. \ref{figure4}, we also plot the value of $r'_{\rm
min}$ for different values of $c_s$ in black lines. As expected,
the black lines are quite close to the corresponding blue ones.

%%%%%%%%%%%%%%%%%%%%%%%%%%%%%%%%%%%%%%%%%  figure 1, figure 1, figure 1%%%%%
%%%%%%%%%%%%%%%%%%%%%%%%%%%%%%%%%%%%%%%%%%%%%

\subsection{Phantom inflation model\label{sec3.4}}

In a spatially flat Friedmann-Lema\^{i}tre-Robertson-Walker
universe, the null energy condition corresponds to the inequality
$\dot H<0$ and hence $n_t<0$. In order to get a blue tilted
gravitational waves, one needs to break the null energy condition,
for example phantom inflation \cite{piao2004} in which the
Lagrangian is given by
\begin{equation}
{\cal L}(\phi)=-{1\over 2}(\partial \phi)^2-V(\phi).
\end{equation}
The tensor-to-scalar ratio $r$ and the tensor spectral
index $n_t$ in this scenario are related to the slow-roll
parameter $\epsilon$ by
\begin{equation}
r=16\epsilon,~~n_t=2\epsilon,
\end{equation}
which leads to the consistency relation for the phantom inflation
\cite{piao2004}
\begin{equation}
n_t=r/8. \label{phan}
\end{equation}
This consistency relation is same as that in (\ref{consistency1}),
except the sign of $n_t$. Therefore the analysis on phantom
inflation is exactly same as that in the canonical single-field
inflation discussed in Sec. \ref{sec3.1}. So, we can easily obtain
the conclusion: $\Delta n_t < |n_t|$ is satisfied only if $r>0.14$
for the EPIC-2m experiment, $r>0.06$ for the ideal CMB experiment.
However, distinguishing phantom inflation from the canonical
single-field slow-roll inflation is a little easier. In order to
attain this goal, we compare $\Delta n_t$ with $\delta n_t=r/4$.
We find that $\Delta n_t<\delta n_t$ is satisfied if $r>0.09$ for
EPIC-2m experiment, and $r>0.06$ for the ideal experiment.

\subsection{Potential-driven G-inflation model \label{sec3.6}}

Recently, an inflation model dubbed as ``G-inflation" was proposed
in \cite{g-inflation0,g-inflation1,g-inflation}. In this class of
models, inflation is driven by a scalar field with Galileon-like
kinetic term. In \cite{g-inflation0}, the authors found that the
model can generate a scale-invariant power spectrum of density
perturbations, and significantly large amplitude of gravitational
waves. The general Lagrangian of G-inflation is of the form
 \bea
 \mathcal{L}_{\phi}=K(\phi,X)-G(\phi,X)\Box\phi, \label{g-inflatioin}
 \ena
where $K$ and $G$ are general functions of $\phi$ and
$X\equiv(\partial\phi)^2/2$. If $G=0$, this model returns to the
general Lorentz-invariant single-field inflation discussed in Sec.
\ref{sec3.2}. In \cite{g-inflation}, the potential-driven
G-inflations were discussed. In this subclass of models, the
Lagrangian in (\ref{g-inflatioin}) has the following form
 \bea
 K(\phi,X)=X-V(\phi),~~G(\phi,X)=g(\phi)X.
 \ena
In the inflationary stage, the energy density is dominated by the
potential $V(\phi)$ under the slow-roll condition. In these
models, the model-independent consistency relation between the
tensor-to-scalar ratio and the tensor spectral index is satisfied
\cite{g-inflation}:
 \bea
 n_t=-\frac{9r}{32\sqrt{6}}, \label{consistency6}
 \ena
which is the smoking-gun evidence for the potential-driven
G-inflation.

Firstly, we investigate whether it is possible to test the
consistency relation in (\ref{consistency6}) by discriminating the
tilted gravitational waves from the scale-invariant one. Very
similar to the case with canonical single-field slow-roll
inflation, we find that $\Delta n_t<|n_t|$ is satisfied if
$r>0.14$ for EPIC-2m mission, and $r>0.06$ for the ideal
experiment.

Another test for the consistency relation is discriminating this
G-inflation from the canonical single-field inflation. To attain
the goal, we compare $\Delta n_t$ with $\delta n_t\equiv
|n_t-r/8|$. Since the two consistency relations
(\ref{consistency1}) and (\ref{consistency6}) are very close with
each other, which induces a very small $\delta n_t=0.01r$, we find
that it is impossible to obtain the condition $\Delta n_t<\delta
n_t$ for EPIC-2m mission so long as $r<1$. Even if we consider the
ideal CMB observations, $\Delta n_t<\delta n_t$ is satisfied only
if $r>0.47$, which is conflicted with the current constraint on
$r$ \cite{komatsu2010}. So we conclude that by the CMB
observations, it is impossible to discriminate the
potential-driven G-inflation from the canonical single-field
inflation by testing the consistency relations.

\section{Conclusions\label{sec4}}

Inflationary scenario has been accepted by most cosmologists,
which naturally solves the flatness, horizon and monopole puzzles
in the standard hot big-bang cosmological model, and predicts the
nearly scale-invariant power spectra of the primordial density
perturbations and gravitational waves. Although the recent
experimental efforts, including the CMB and the large-scale
structure, have led to a robust detection of the primordial
density perturbations, and indirectly supported the existence of
the early inflationary stage, how to distinguish different
inflation models still remains an outstanding experimental
challenge. It depends on how accurately we can measure the
primordial perturbations, in particular the primordial
gravitational waves.

In this paper, based on the potential future CMB observations by
the planned CMBPol mission and an ideal CMB experiment, we
investigated the possible tests for the consistency relations in
several classes of single-field inflation models: the canonical
single-field slow-roll inflation, the general Lorentz-invariant
single-field inflation, the phantom inflation and the
potential-driven G-inflation. For the canonical single-field
inflation, phantom inflation and the potential-driven G-inflation,
we found that the consistency relations are quite hard to be
tested, due to the smallness of spectral index $n_t$. For example,
the testing is possible only if $r>0.14$ for CMBPol mission and
$r>0.06$ for the ideal experiment.

Dramatically the situation becomes quite promising for the general
Lorentz-invariant single-field inflation with large non-local
non-Gaussianity,
%the two-field inflation with strong correlation
%between adiabatic and isocurvature perturbations,
%and the spacetime non-commutative inflation with low
%non-commutative scale
because the value of $|n_t|$ could be quite
large compared to that in the canonical single-field inflation for
a given $r$ in these cases. For the general Lorentz-invariant
single-field inflation with the non-Gaussian parameters $f_{\rm
NL}^{\rm equil}=f_{\rm NL}^{\rm orth}=-20$, the testing is
possible if $r>0.015$ for CMBPol mission and $r>0.006$ for the
ideal experiment.
%Similarly, for the spacetime non-commutative
%inflation with $\mu\rightarrow1$, the test is possible if $r>0.04$
%for CMBPol mission and $r>0.01$ for the ideal experiment.

In the end of this paper, it is worthy to point out that the
similar analysis can be applied to observationally test some other
inflationary scenarios \cite{chain,wands2002}. At the same time,
in addition to the relations related to $r$ and $n_t$, the other
inflationary consistency relations (see
\cite{other1,other2,other3} for instance) can also be used to
distinguish different inflation models. We leave it as a future
work.

%%%%%%%%%%%%%%%%%%%%%%%%%%%%%%%%%%%%%%%%%%%%%%%%%%%%%%%%%%%%%%%%%%%%%%%%%%%%%%%%%%%%%
%%%%%%%%%%%%%%%%%%%%%%%   Appendix   %%%%%%%%%%%%%%%%%%%%%%%%%%%%%%%%
%%%%%%%%%%%%%%%%%%%%%%%%%%%%%%%%%%%%%%%%%%%%%%%%%%%%%%%%%%%%%%%%%%%%%%%%%%%%%%%%%%%%%

%%%%%%%%%%%%%%%%%%%%%%%%%%%%%%%%%%%%%%%%%%%%%%%%%%%%%%%%%%%%%%%%%%%%%%%%%%%%%%%%%%%%%
%%%%%%%%%%%%%%%%%%%%%%%   Acknowledge   %%%%%%%%%%%%%%%%%%%%%%%%%%%%%%%%
%%%%%%%%%%%%%%%%%%%%%%%%%%%%%%%%%%%%%%%%%%%%%%%%%%%%%%%%%%%%%%%%%%%%%%%%%%%%%%%%%%%%%

\section*{Acknowledgements}
We are very grateful to H.Noh, Y.S.Piao and A.Vikman for
invaluable discussions. W.Z. is partially supported by NSFC Grants
Nos. 11173021 and 11075141. Q.G.H. is supported by the project of
Knowledge Innovation Program of Chinese Academy of Science and a
NSFC Grant No. 10975167.


\begin{thebibliography}{35}

\bibitem{weinberg2008}
S. Weinberg, ~ (Oxford University Press, New York, 2008).

\bibitem{inflation}
A. H. Guth, ~\prd {\bf 23}, 347 (1981); D. H. Lyth and A. Riotto,
~\pr {\bf 314}, 1 (1999).

\bibitem{zeldovich1981}
Ya. B. Zeldovich,~{Pis'ma Astron. Zh}~{\bf 7}, 579 (1981); L. P.
Grishchuk and Ya. B. Zeldovich, in {\it Quantum Structure of Space
and Time}, Eds. M. Duff and C. Isham, (Cambridge University Press,
Cambridge, England, 1982), p. 409; Ya. B. Zeldovich, {\it
Cosmological field theory for observational astronomers}, {\em
Sov. Sci. Rev. E Astrophys. Space Phys.,} Harwood Academic
Publishers, Vol. 5, pp. 1-37 (1986) ({\em
http://nedwww.ipac.caltech.edu/level5/Zeldovich
/Zel{\_}contents.html}); A. Vilenkin, in ``{\em The Future of
Theoretical Physics and Cosmology}", Eds. G.W.Gibbons, E. P. S.
Shellard and S. J. Rankin (Cambridge University Press, Cambridge,
England, 2003); L. P. Grishchuk, ~{Space Science Reviews} {\bf
148}, 315 (2009) [arXiv:0903.4395].

\bibitem{grishchuk1974}
L. P. Grishchuk,~{Sov. Phys. JETP} {\bf 40}, 409 (1975); ~ {Ann.
N. Y. Acad. Sci.}~{\bf 302}, 439 (1977); ~{JETP Lett.}~{\bf 23},
293 (1976)
(\emph{http://www.jetpletters.ac.ru/ps/1801/article{\_}27514.pdf});
~ {Sov. Phys. Usp.}~{\bf 20}, 319 (1977); ~in \emph{General
Relativity and John Archibald Wheeler}, Eds. I. Ciufolini and R.
Matzner, (Springer, New York, 2010) pp. 151-199 [arXiv:0707.3319].

\bibitem{mukhanov1991}
V. F. Mukhanov, H. A. Feldman, and R. H. Brandenberger, ~\pr~{\bf
215}, 203 (1992).


\bibitem{grishchuk2001}
L. P. Grishchuk, ~\lnp {\bf 562}, 167 (2001); Y. Zhang, Y. F.
Yuan, W. Zhao and Y. T. Chen, ~\cqg {\bf 22}, 1383 (2005); Y.
Watanabe and E. Komatst,  \prd {\bf 73}, 123515 (2006); W. Zhao
and Y. Zhang, ~\prd {\bf 74}, 043503 (2006); L. A. Boyle and P. J.
Steinhardt, ~\prd {\bf 77}, 063504 (2008); M. Giovannini, ~\cqg
{\bf 26}, 045004 (2009); M. L. Tong and Y. Zhang, ~\prd {\bf 80},
084022 (2009).




\bibitem{knox1999}
L. Knox, ~\prd {\bf 60}, 103516 (1999).

\bibitem{wmap_notation}
H. V. Peiris {et al.,} ~\apjs {\bf 148}, 213 (2003).


\bibitem{efstathiou2006}
S. Chongchitnan and G. Efstathiou, \prd {\bf 73}, 083511 (2006).


\bibitem{verde2006}
L. Verde, H. Peris and R. Jimenez, ~\jcap {\bf 0601}, 019 (2006).



\bibitem{cmbpol}
D. Baumann et al., ~AIP Conf. Proc. {\bf 1141} 10 (2009).


\bibitem{peiris2011}
M. J. Mortonson, H. V. Peiris and  R. Easther,  \prd {83}, 043505
(2011).


\bibitem{easson2010}
D. A. Easson and B. Powell, ~arXiv:1009.3741; D. A. Easson and B.
Powell, ~\prd {\bf 83}, 043502 (2011).









\bibitem{polnarev1985}
A. Polnarev, ~Sov. Astron. {\bf 29}, 607 (1985); L. P. Grishchuk,
~\prl {\bf 70}, 2371 (1993); U. Seljak and M. Zaldarriaga, ~\prl
{\bf 78}, 2054 (1997); M. Kamionkowski, A. Kosowsky and A.
Stebbins, ~\prl {\bf 78}, 2058 (1997); J. R. Printchard and M.
Kamionkowski, ~Ann. Phys. (N.Y.) {\bf 318}, 2 (2005); W. Zhao and
Y. Zhang, ~\prd {\bf 74}, 083006 (2006); D. Baskaran, L. P.
Grishchuk and A. G. Polnarev, ~\prd {\bf 74}, 083008 (2006); T. Y.
Xia and Y. Zhang, ~\prd {\bf 79} 083002 (2009).



\bibitem{komatsu2010}
E. Komatsu et al., ~\apjs {\bf 192}, 18 (2011).

\bibitem{zbg2009}
W. Zhao, D. Baskaran and L. P. Grishchuk, ~\prd {\bf 79}, 023002
(2009); ~\prd {\bf 80}, 083005 (2009); ~\prd {\bf 82}, 043003
(2010); W. Zhao, ~\prd {\bf 79}, 063003 (2009); W. Zhao and L. P.
Grishchuk, ~\prd {\bf 82}, 123008 (2010).

\bibitem{quad}
P. Ade et al. (QUaD Collaboration), ~\apj {\bf 674}, 22 (2008); C.
Pryke et al. (QUaD Collaboration), ~\apj {\bf 692}, 1247 (2009);
S. Gupta et al. (QUaD Collaboration), ~ \apj {\bf 716}, 1040
(2010).

\bibitem{bicep}
H. C. Chiang et al., ~ \apj {\bf 711}, 1123 (2010).


\bibitem{currentquiet}
C. Bischoff et al. (QUIET Collaboration), ~arXiv:1012.3191.

\bibitem{clover}
C. E.~North {et al.,} ~arXiv:0805.3690.


\bibitem{polarbear}
http://bolo.berkeley.edu/POLARBEAR/index.html.



\bibitem{quiet}
D.~Samtleben, ~arXiv:0802.2657; http://quiet.uchicago.edu/.

\bibitem{quijote}
J. A. Rubino-Martin et al., ~arXiv:0810.3141;
http://www.iac.es/project/cmb/quijote/.

\bibitem{sptpol}
http://pole.uchicago.edu/.



\bibitem{actpol}
http://wwwphy.princeton.edu/act/.


\bibitem{qubic}
J. Kaplan, ~arXiv:0910.0391; E. Battistelli et al. (QUBIC
Collaboration), ~arXiv:1010.0645.

\bibitem{ebex}
http://groups.physics.umn.edu/cosmology/ebex/index.html.

\bibitem{pappa}
A. Kogut, et al., ~New Astron. Rev. {\bf 50}, 1009 (2006).

\bibitem{spider}
B. P. Crill, et al., ~arXiv:0807.1548.



\bibitem{taskforce}
J. Bock et al.,  astro-ph/0604101.



\bibitem{planck}
{Planck} Collaboration, astro-ph/0604069.


\bibitem{b-pol}
B-Pol Collaboration, ~Exper. Astron. {\bf 23}, 5 (2009);
http://www.b-pol.org/index.php.


\bibitem{litebird}
http://cmbpol.kek.jp/litebird/.





\bibitem{core}
COrE Collaboration, ~arXiv:1102.2181;
http://www.core-mission.org/.

\bibitem{zhao2009a}
W. Zhao and D. Baskaran, ~\prd {\bf 79}, 083003 (2009).

\bibitem{zhao2009b}
W. Zhao and W. Zhang, ~\plb {\bf 677}, 16 (2009).

\bibitem{ma2010}
Y. Z. Ma, W. Zhao and M. L. Brown, ~\jcap {\bf 1010}, 007 (2010).

\bibitem{zhao2011}
W. Zhao, ~\jcap {\bf 1103}, 007 (2011).


\bibitem{GrishchukSolokhin}
L.~P.~Grishchuk and M.~Solokhin, ~\prd {\bf 43}, 2566 (1991).


\bibitem{turner}
A. Kosowsky and M. S. Turner, ~\prd {\bf 52}, 1739 (1995).




\bibitem{cosmomc}
A. Lewis and S. L. Bridle, ~\prd {\bf 66}, 103511 (2002).




\bibitem{efstathiou2009}
G. Efstathiou and S. Gratton, ~\jcap {\bf 0906}, 011 (2009).

\bibitem{cmbpol2}
J. Bock  et al. (EPIC Collaboration), ~arXiv:0805.4207; J. Bock
et al. (EPIC Collaboration), ~arXiv:0906.1188.

\bibitem{camb}
http://camb.info/.

\bibitem{ebmixture}
A. Lewis, ~\prd {\bf 68}, 083509 (2003); A. Lewis, A. Challinor
and N. Turok, ~\prd {\bf 65}, 023505 (2001); E. F. Bunn, M.
Zaldarriaga, M. Tegmark and A. de Oliveira-Costa, ~\prd {\bf 67},
023501 (2003); K. M. Smith, ~\prd {\bf 74}, 083002 (2006); K. M.
Smith and M. Zaldarriaga, ~\prd {\bf 76}, 043001 (2007); W. Zhao
and D. Baskaran, ~\prd {\bf 82}, 023001 (2010); J. Kim and P.
Naselsky, ~\aanda {\bf 519}, A104 (2010); J. Kim,
~arXiv:1010.2636; E. F. Bunn, ~arXiv:1008.0827; J. Bowyer, A. H.
Jaffe and D. I. Novikov, \~arXiv:1101.0520.




\bibitem{lensingreview}
A. Lewis and A. Challinor, ~\pr {\bf 429}, 1 (2006).



\bibitem{song2002}
L. Knox and Y. S. Song, ~\prl {\bf 89}, 011303 (2002);  M. Kesden,
A. Cooray and M. Kamionkowski, \prl {\bf 89}, 011304 (2002).



\bibitem{lensing1}
W. Hu and T. Okamoto, ~\apj {\bf 574}, 566 (2002); T. Okamoto and
W. Hu, ~\prd {\bf 67}, 083002 (2003).




\bibitem{lensing2}
C. M. Hirata and U. Seljak, ~\prd {\bf 68}, 083002 (2003); U.
Seljak and C. M. Hirata, ~\prd {\bf 69}, 043005 (2004).



\bibitem{Alabidi:2008ej}
  L.~Alabidi and J.~E.~Lidsey,
  Phys.\ Rev.\  D {\bf 78}, 103519 (2008);L.~Alabidi and I.~Huston,
  JCAP {\bf 1008}, 037 (2010);Y.~Z.~Ma and X.~Zhang,
  JCAP {\bf 0903}, 006 (2009).



\bibitem{transplanck}
A. Ashoorioon, J. L. Hovdebo and R. B. Mann, ~\npb {\bf 727}, 63
(2005).



\bibitem{smith2006}
Y. S. Song and L. Knox, ~\prd {\bf 68}, 043518 (2003); T. L.
Smith, H. V. Peiris and A. Cooray, ~\prd {\bf 73}, 123503 (2006).


\bibitem{liddle1983}
A. D. Linde, ~\plb {\bf129}, 177 (1983).


\bibitem{kinflation}
C. Armendariz-Picon, T. Damour and V. F. Mukhanov, ~\plb {\bf
458}, 209 (1999); J. Garriga and V. F. Mukhanov, ~\plb {\bf 458},
219 (1999); E. Babichev, V. Mukhanov and A. Vikman, ~\jhep {\bf
0802}, 101 (2008); V. Mukhanov and A. Vikman, ~\jcap {\bf 0602},
004 (2006).



\bibitem{alishahiha2004}
M. Alishahiha, E. Silverstein and D. Tong, ~\prd {\bf 70}, 123505
(2004); E. Silverstein and D. Tong, ~\prd {\bf 70}, 103505 (2004).


\bibitem{seery2005}
D. Seery and J. E. Lidsey, ~\jcap {\bf 0506}, 003 (2005).

\bibitem{panotopoulos2007}
G. Panotopoulos, ~\prd {\bf 76}, 127302 (2007).





\bibitem{chen2007}
X. Chen, M. X. Huang, S. Kachru and G. Shiu, ~\jcap {\bf 0701},
002 (2007).


\bibitem{komatsu2000}
E. Komatsu and D. N. Spergel, ~\prd {\bf 63}, 063002 (2001).

\bibitem{Babich2004}
D. Babich, P. Creminelli and M. Zaldarriaga, ~\jcap {\bf 0408},
009 (2004).

\bibitem{Senatore2009}
L. Senatore, K. M. Smith and M. Zaldarriaga, ~\jcap {\bf 1001},
028 (2010).

\bibitem{other2}
Q. G. Huang, ~\jcap {\bf 1005}, 016 (2010).





\bibitem{zaldarriaga2004}
D. Babich and M. Zaldarriaga,  \prd {\bf 70}, 083005 (2004).

\bibitem{yadav2007}
A. P. S. Yadav, E. Komatsu and B. D. Wandelt,  \apj {\bf 664}, 680
(2007).

\bibitem{smith2009a}
K. M. Smith, L. Senatore and M. Zaldarriaga,  \jcap {\bf 0909},
006 (2009).

\bibitem{komatsu}
E.~Komatsu, private communication.








\bibitem{piao2004}
Y. S. Piao and Y. Z. Zhang, ~\prd {\bf 70}, 063513 (2004).






\bibitem{g-inflation0}
T. Kobayashi, M. Yamaguchi and J. Yokoyama, ~\prl {\bf 105},
231302 (2010).


\bibitem{g-inflation1}
S. Mizuno and K. Koyama, ~\prd {\bf 82}, 103518 (2010); C.
Burrage, C. de Rham, D. Seery and A. J. Tolley, ~\jcap {1101}, 014
(2011); P. Creminelli, G. DAmico, M. Musso, J. Norena and E.
Trincherini, ~\jcap {\bf 1102}, 006 (2011); S. Renaux-Petel,
arXiv:1105.6366; S. Renaux-Petel, arXiv:1107.5020; S.
Renaux-Petel, S. Mizuno and K. Koyama, arXiv:1108.0305.




\bibitem{g-inflation}
K. Kamada, T. Kobayashi, M. Yamaguchi and J. Yokoyama, ~ \prd {\bf
83}, 083515 (2011).

\bibitem{chain}
A. Ashoorioon, ~\jcap {\bf 1004}, 002 (2010).


\bibitem{wands2002}
D. Wands, N. Bartolo, S. Matarrese and A. Riotto, ~\prd {\bf 66},
043520 (2002);

\bibitem{other1}
P. Creminelli and M. Zaldarriaga, ~\jcap {\bf 0401}, 006 (2004).

\bibitem{other3}
C. Cheung, A. L. Fitzpatrick, J. Kaplan and L. Senatore,~\jcap
{\bf 0802}, 021 (2008).






\end{thebibliography}
\end{document}